\documentclass[twocolumn,prc,showpacs,showkeys,preprintnumbers,unsortedaddress,amsmath,amssymb,floatfix]{revtex4}

\usepackage{graphicx}
\usepackage{bm}

\begin{document}

\title{From Brueckner Approach to Skyrme-type Energy Density Functional}

\author{L. G. Cao$^{1}$, U. Lombardo$^{1,2}$, C. W. Shen$^{3}$, Nguyen Van Giai$^{4}$}

\affiliation{$^{1}$LNS-INFN, Via Santa Sofia 44, I-95123 Catania,
Italy}

\affiliation{$^{2}$Dipartimento di Fisica dell'Universit\`a, Viale
Andrea Doria 6, I-95123 Catania, Italy}

\affiliation{$^{3}$School of Science, Huzhou Teachers College,
No.1 Xueshi Road, Huzhou 313000, P.R. of China }

\affiliation{$^{4}$Institut de Physique Nucl\'eaire, Universit\'e
Paris Sud F-91406 Orsay CEDEX, France}

\date{ \today}

\begin{abstract}

A Skyrme-like effective interaction is  built up from the equation
of state of nuclear matter. The latter is calculated  in the
framework of the Brueckner-Hartree-Fock approximation with two and
three body forces. A complete Skyrme parametrization requires a fit
of the neutron and proton effective masses and also the Landau
parameters. The new parametrization is probed on the properties of a
set of closed-shell and closed-subshell nuclei, including binding
energies and charge radii.

\end{abstract}
\pacs{21.30.Fe, 21.60.Jz, 21.65.+f, 21.10.Dr}

 \keywords{effective
nuclear interactions, nuclear matter, Hartree-Fock approximation,
Brueckner-Hartree-Fock}

\maketitle

\section{Introduction}
For more than 30 years the microscopic description of nuclear ground
state properties is relying on the self-consistent mean field
approach, or Hartree-Fock (HF) approach built from effective
in-medium nucleon-nucleon interactions\cite{vau72}. The most popular
interaction is the Skyrme-type interaction\cite{Skyrme} whose
analytic form leads to considerable simplifications of the HF
calculations in finite nuclei. The general point of view is that the
Skyrme interaction is a phenomenological one whose parameters are
directly adjusted on a few selected observables taken from infinite
matter and some doubly-magic nuclei. In its most sophisticated
versions, the Skyrme-type force can predict binding energies of all
measured nuclei with an overall error of less than 0.7
MeV\cite{Goriely}. Other parametrizations try to improve the
description of systems with a large neutron excess\cite{Chabanat} by
incorporating constraints from variational calculations performed
for neutron-rich and pure neutron matter\cite{wiringa,akmal}.

On the other hand, important progress have been made in the recent
years concerning Brueckner-Hartree-Fock (BHF) calculations of
infinite matter\cite{song,tbf}. For many years, a major drawback
of the BHF approach was that the empirical saturation point of
symmetric nuclear matter could not be reproduced if one starts
with only a realistic two-body bare interaction (the so-called
Coester line problem). This is the reason why in the early
attempts to derive from a Brueckner G-matrix an effective
interaction suitable for HF calculations of finite nuclei it was
necessary to renormalize phenomenologically the
G-matrix\cite{Negele, Negele72}. Two recent achievements have led
the BHF approach to a quite satisfactory status: one is that the
convergence of the hole-line expansion has been proved to occur
already at the level of three-body correlations\cite{song}; the
other one is that the inclusion of three-body forces improves to a
large extent the prediction of the saturation point\cite{tbf}.

Thus, it is timely to re-examine how one can relate the BHF
description of homogeneous matter to a HF description of finite
nuclei. Our strategy is to look for a Skyrme-like parametrization
adjusted so as to reproduce the BHF results calculated in
symmetric nuclear matter and also spin- and isospin polarized
matter, then to examine the HF predictions of this parametrization
in nuclei. In our approach it is possible to determine all the
parameters of the force except the two-body spin-orbit parameter
$W_0$ because this component of the force gives no contribution in
homogeneous matter. This will remain an adjustable parameter when
calculating finite nuclei, and its value is fixed on the $1p1/2 -
1p3/2$ splitting in $^{16}$O as it is usually done. Of course, the
two-body Coulomb interaction has also to be added.

In a recent work, Baldo et al.\cite{baldo0} have studied the
relation between the Brueckner results in infinite matter and some
of the Skyrme parameters, but the number of constraints used was
insufficient to determine the velocity-dependent part of the Skyrme
force, which is a very important part since it governs the effective
mass behaviour.

In this work, we start from the recent BHF calculations of
Refs.\cite{ebhf,tbf} extended to spin- and isospin polarized
homogeneous matter. The paper is organized as follows: in Sec.II a
brief review of relevant BHF predictions is given. In Sec.III we
explain how the Skyrme parameters are determined. Sec.IV presents
and discusses the HF results obtained for selected closed-shell
nuclei. Concluding remarks are given in Sec.V.

\section{Review of the relevant BHF predictions}

For densities around the saturation value and smaller, the BHF
approximation, in which the energy shift is truncated at two
hole-line level, is a quite good approximation. In that region in
fact we are allowed to neglect the three-body correlations, which
anyhow give a small contribution when adopting the continuous
choice for the auxiliary potential\cite{song}. On the other hand,
the BHF approximation embodies both the inert core pure BHF mean
field and the core polarization term, the latter arising as
rearrangement potential\cite{ebhf}. Thus, the main effects of the
correlations on the effective mass will be taken into account. The
fit of the microscopic effective mass is a qualifying aspect of
the present parametrization, since this quantity is one of the
least well constrained properties of the phenomenological Skyrme
forces\cite{stone}.

Let us briefly review the salient features of the BHF equation of
state (EoS):

{\it i) saturation point}

 The BHF approach with only two-body
force sizeably overestimates the saturation density\cite{baldo}.
This is commonly attributed to the missing effect of the three body
force in the region above the empirical saturation density. After
introducing a three body force in the calculations the saturation
density is improved without appreciable change of the corresponding
energy. In the most recent Brueckner calculations\cite{tbf} it was
estimated , $\rho_0\approx 0.18 fm^{-3}$, still beyond the range of
the empirical values of the central density of nuclei, and the
energy $\frac{E}{A}\approx -14.8 MeV$. The main effect of such a
failure is expected to appear in the
calculation of the neutron and proton density profiles.\\
{\it ii) effective mass}\\
The nucleon effective mass inside the medium is an outcome of the
BHF approach and as such, it must be reproduced by the equivalent
Skyrme parametrization. It is related to the momentum dependence
of the on-shell self-energy $\Sigma(\epsilon_k,k)$, where
$\epsilon_k$ is the quasiparticle energy. It plays an important
role not only in the theoretical description of the transport
phenomena, including heavy ion collision (HIC) simulations, but
also for the level density of nuclei. Recently the isospin
splitting of the effective mass has been the subject of some
debate since different approaches predict contradictory results
\cite{split,samma,ma,Toro}. In the framework of the Brueckner
approach it is possible to trace the isospin splitting of the
effective mass back to properties  of the bare nuclear interaction
\cite{split,bomb}. It turns out that the neutron effective mass is
linearly increasing with the neutron excess while the proton
effective mass is symmetrically decreasing. Embodying the neutron
and proton effective masses in the fit one may expect important
isospin effects in observables which are sensitive to the
effective mass itself.

iii){\it symmetry energy}

 The symmetry energy $a_s(\rho)$ has
stimulated a lot of interest for its relevance in HIC physics,
nuclear astrophysics and exotic nuclei. In fact it is related to
the isospin splitting of effective mass, neutron and proton mean
fields, etc... In particular, the neutron skin in neutron-rich
nuclei seems to be very sensitive to the details of the density
dependence of $a_s(\rho)$\cite{brown}. In the microscopic
approaches, the saturation point becomes less and less stable with
increasing neutron excess and at some critical point, before
reaching the conditions of pure neutron matter, it disappears.
This transition formally amounts to the transition from a minimum
to an inflexion point in the function $\frac{E}{A}(\beta,\rho)$.
Correspondingly the symmetry energy would also exhibit an
inflexion point as a function of density. This entails that any
parametrization of $a_s(\rho)$ such as $\rho^\alpha$, which is
often adopted in calculations, is not suitable to reproduce this
behavior. This seems to be the case with the Skyrme forces. In the
BHF approximation the symmetry energy at the saturation point
turns out to be $a_s(\rho_0)\approx 34. MeV$ \cite{tbf}.
 At low density it is
independent of the force \cite{baldo0} and therefore it can be
considered well established from a quantum-mechanical many body
theory. Above the saturation point the three body force has a strong
influence on the symmetry energy. The BHF prediction is in a rather
good agreement with the relativistic Dirac-Brueckner\cite{samma} but
both diverge from variational calculations\cite{wiringa,akmal}and
also from some Skyrme forces (for a discussion, see
Ref.\cite{stone}). The structure of neutron stars has been addressed
as a possible constraint for the EoS of asymmetric nuclear matter
and, in particular, for the symmetry energy, but so far the
calculations do not give a definite answer.\\
iv) {\it Landau parameters}\\
 In order to have a full determination
of the Skyrme-force parameters and not only some combinations of
them, it is not sufficient to fit bulk properties such as binding
energies, effective masses and symmetry energy, as we shall see in
Sec.III. We will use the additional constraint of reproducing the
$G_0$ Landau parameter extracted by extending the BHF calculations
to spin and isospin asymmetric nuclear matter\cite{zuo2003}.
 The Brueckner predictions for the Landau parameters\cite{zuo2003} have
proved to reproduce the existing experimental data; in particular
the parameter $G_0'$ is consistent with the centroid energy of the
Gamow-Teller giant resonance\cite{suz}. So far only the values of
the Landau parameters at the saturation point can be tested. At
lower density experimental information can come from the study of
giant resonances in exotic nuclei.

\section{Determination of Skyrme force parameters}

The standard form of the Skyrme effective interaction is:
\begin{eqnarray}
&V(\mathbf{r}_1,\mathbf{r}_2)=&t_0(1+x_0P_{\sigma})\delta(\mathbf{r}) \nonumber \\
&&+\frac{1}{2}t_1(1+x_1P_{\sigma})[\mathbf{P}'^2\delta(\mathbf{r})+\delta(\mathbf{r})\mathbf{P}^2] \nonumber \\
&&+t_2(1+x_2P_{\sigma})\mathbf{P}'\cdot\delta(\mathbf{r})\mathbf{P}
\nonumber \\
&&+\frac{1}{6}t_3(1+x_3P_{\sigma})[\rho(\mathbf{R})]^{\sigma}\delta(\mathbf{r})\nonumber
\\
&&+i W_0\bf{\sigma} \cdot
[\mathbf{P}'\times\delta(\mathbf{r})\mathbf{P}] ~, \label{eq1}
\end{eqnarray}
 where $\mathbf{r}=\mathbf{r}_1-\mathbf{r}_2$ is the relative coordinate of
the two particles and $\mathbf{R}=(\mathbf{r}_1+\mathbf{r}_2)/2$
the center of mass coordinate. $\mathbf{P}=(\nabla_1-\nabla_2)/2i$
is the relative momentum acting on the right, and $\mathbf{P}'$
its conjugate acting on the left.
$P_\sigma=(1+\sigma_1\cdot\sigma_2)/2$ is the spin-exchange
operator.

The force parameters are the $t_i$ and $x_i$, the power $\sigma$
of the density dependence, and the spin-orbit strength $W_0$. As
already mentioned, the latter parameter will be adjusted
phenomenologically in some specific nucleus. As for the $\sigma$
parameter, it is difficult to extract it in a unique way just from
nuclear matter bulk properties. In the literature there are
several classes of Skyrme forces characterized by the value of
$\sigma$, the most common values being 1/6, 1/3, 1. Since we do
not have enough constraints from BHF calculations of nuclear
matter to determine all the free parameters, we choose to adopt in
this work $\sigma$=1/6 and to concentrate on finding the remaining
parameters.

The basic inputs of this work are the results of the BHF
self-energy which includes the core polarization term (called in
the literature extended BHF approximation (EBHF)\cite{ebhf}) and
the EOS with two and three-body forces\cite{tbf}. The procedure to
determine the force parameters proceeds through three main steps.
The first step concerns the fit of the nucleon effective mass in
symmetric and non-symmetric nuclear matter. This enables one to
find the values of two important combinations - $\Theta_s$ and
$\Theta_v$ - of $t_1,t_2,x_1,x_2$. In the second step we look at
the energy per particle in symmetric and non-symmetric nuclear
matter as a function of total density and neutron-proton
asymmetry. The fit of these quantities determines several families
of ($t_0,t_3,x_0$,$x_3$) parameters. In the last step, we make use
of the constraints imposed by the value of the $G_0$ Landau
parameter and by a combination of parameters governing the surface
properties of finite nuclei. In this way, the remaining parameters
$t_1,t_2,x_1,x_2$ are uniquely determined for each parameter
family since we already know the values of $\Theta_s$ and
$\Theta_v$. Because the fit of BHF bulk properties has been
supplemented with the surface condition one may hope that the
parameter sets thus obtained could describe reasonably well finite
nuclei.

\subsection {Effective masses}

In symmetric nuclear matter, the isoscalar effective mass of a
nucleon has the following expression (here and in the rest of this
paper we follow the same notations as in Ref.\cite{Chabanat}):
\begin{equation}
\frac{m^*_s}{m}=(1+\frac{m}{8\hbar^2}\rho\Theta_s)^{-1} ~,
\label{eq3}
\end{equation} where $\Theta_s=[3t_1+(5+4x_2)t_2]$.
One can also define an isovector effective mass as \begin{equation}
\frac{m^*_v}{m}=(1+\frac{m}{4\hbar^2}\rho\Theta_v)^{-1}~,
\label{eq55}
\end{equation}
where $\Theta_v=t_1(x_1+2)+t_2(x_2+2)$. In asymmetric nuclear matter
with an asymmetry parameter $\beta = (N-Z)/A$, the nucleon effective
mass is
\begin{equation}
\frac{m^*_q}{m}=(1+\frac{m}{8\hbar^2}\rho\Theta_s-\frac{m}{8\hbar^2}q(2\Theta_v-\Theta_s)\beta\rho)^{-1}
~, \label{eq5}
\end{equation}
with $q=1$ for neutrons and $q=-1$ for protons.

To obtain $\Theta_s$ we fit the values of the nucleon effective
mass calculated in symmetric nuclear matter in EBHF
approximation\cite{ebhf} with the three-body force effects
included. The fit is illustrated in Fig.1, and the resulting value
from the fit is $\Theta_s = 400.8$ MeV fm$^5$. Next, we can
determine $\Theta_v$ by fitting $m^*_q/m$ calculated in asymmetric
nuclear matter. The fits are shown in Fig.2, and the corresponding
value of $\Theta_v$ is $356.4$ MeV fm$^5$. Actually, the fits of
$m^*_q/m$ remain good beyond $\beta$=0.4~.

\begin{figure}[hbtp]
\includegraphics[scale=0.3]{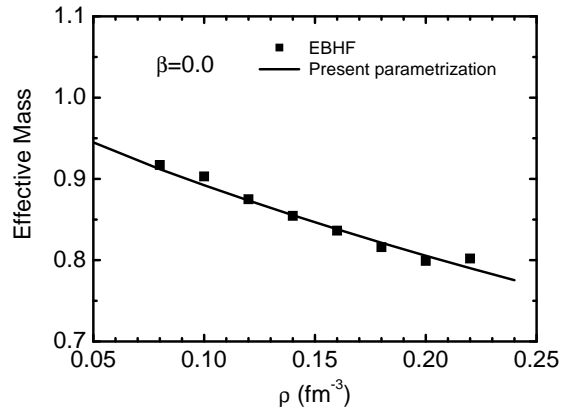}
\vglue -3.cm \caption{Nucleon effective mass $m^*/m$ in symmetric
nuclear matter. } \label{Fig.1}
\end{figure}

At this point we can already make an interesting observation. Some
Skyrme parametrizations of the literature predict that, for a
fixed density, the proton and neutron effective masses are
respectively an increasing and decreasing function of the
asymmetry parameter $\beta$\cite{stone}. This behaviour is
opposite to that predicted by EBHF calculations\cite{split}. In
Fig.3 we show a comparison of effective masses calculated with the
parameter set SLy4\cite{Chabanat} and with our values of
$\Theta_s$ and $\Theta_v$. Thus, our Skyrme parametrization will
differ from some usual parametrizations as far as the isospin
splitting of effective masses is concerned. Recently, it was
pointed out\cite{ma} that effective masses obtained in
Dirac-Brueckner-Hartree-Fock have a similar behaviour to that of
EBHF, and opposite to that of the relativistic mean field (RMF)
model. It must also be noted that, if one includes the Fock terms
in a relativistic Hartree-Fock description, the trend of the
neutron-proton mass splitting becomes closer to that of the
Dirac-Brueckner-Hartree-Fock and therefore, the behaviour of the
RMF mass is due to the omission of exchange terms\cite{Long2005}.

\begin{figure}[hbtp]

\includegraphics[scale=0.3]{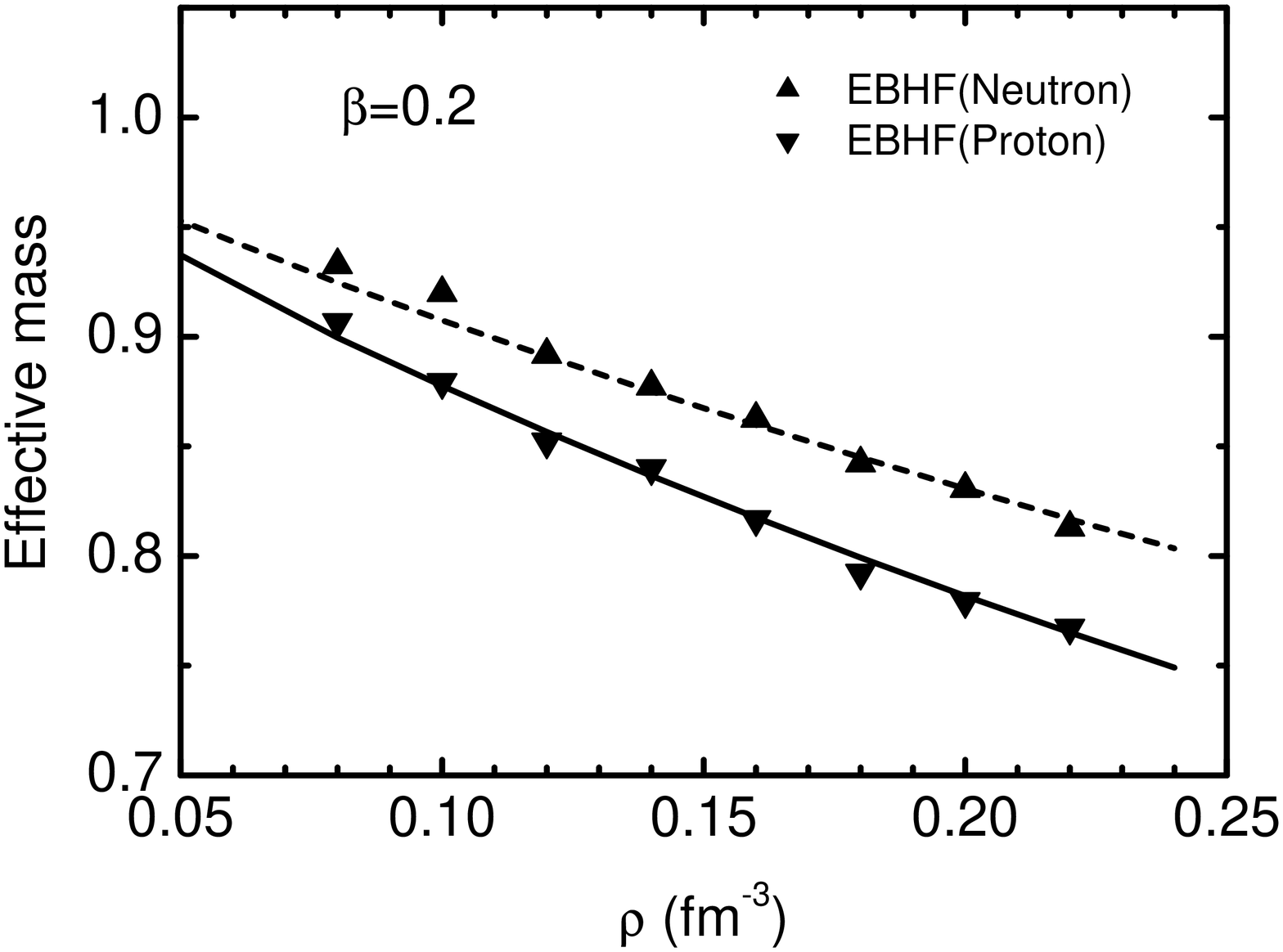}
\vglue -3.3cm
\includegraphics[scale=0.3]{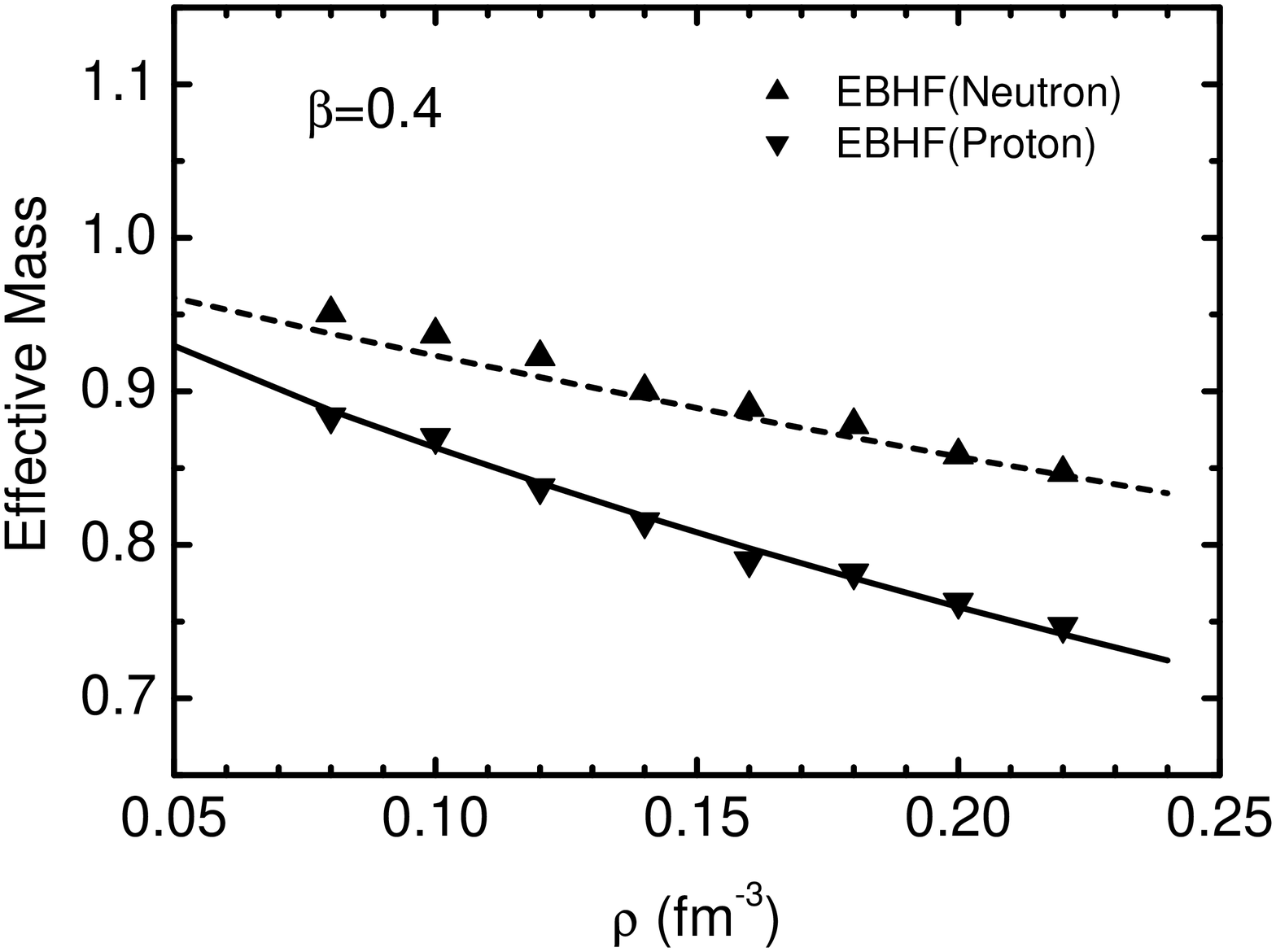}
\vglue -3.cm \caption{Nucleon effective masses $m^*_q/m$ in
asymmetric nuclear matter, at two different asymmetries. Triangles
are EBHF results, the lines are the fits. } \label{Fig.2}
\end{figure}

\begin{figure}[t]
\vglue -0.8cm
\includegraphics[scale=0.3]{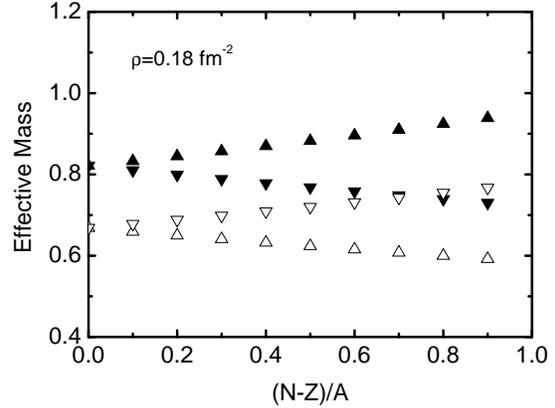}
\vglue -3.cm \caption{Neutron (upward triangles) and proton
(downward triangles) effective masses $m^*/m$ calculated with SLy4
(in white) and with the present parametrization (in black).}
\label{Fig.3}
\end{figure}

\subsection {Energy per particle}

In symmetric matter the energy per particle has the following
expression:
\begin{equation}
\frac{E}{A}(\rho)=\frac{3\hbar^2}{10m}(\frac{3\pi^2}{2})^{\frac{2}{3}}\rho^{\frac{2}{3}}+\frac{3}{8}t_0\rho
+\frac{3}{80}\Theta_s(\frac{3\pi^2}{2})^{\frac{2}{3}}\rho^{\frac{5}{3}}+\frac{1}{16}t_3\rho^{\sigma+1}
~. \label{eq2}
\end{equation} More generally, in asymmetric matter characterized by
an asymmetry parameter $\beta$ the energy per particle is:
\begin{eqnarray}
&\frac{E}{A}(\beta,\rho)=&\frac{3\hbar^2}{10m}(\frac{3\pi^2}{2})^{\frac{2}{3}}\rho^{\frac{2}{3}}F_{\frac{5}{3}}+\frac{1}{8}t_0\rho
[3-(2x_0+1)\beta^2] \nonumber \\
&&+\frac{1}{48}t_3\rho^{\sigma+1}[3-(2x_3+1)\beta^2] \nonumber\\
&&+\frac{3}{40}(\frac{3\pi^2}{2})^{\frac{2}{3}}\rho^{\frac{5}{3}}
[\Theta_vF_\frac{5}{3}+\frac{1}{2}(\Theta_s-2\Theta_v)F_\frac{8}{3}]
~,\nonumber \\ \label{eq4}
\end{eqnarray}
where $F_m(\beta)=\frac{1}{2}[(1+\beta)^m+(1-\beta)^m]$.

We first start with the case of symmetric matter. The fit of the
BHF values of $\frac{E}{A}(\beta=0,\rho)$ determines several
possible values for the couple ($t_0,t_3$). To limit the number of
possible couples we impose that the compression modulus $K_\infty$
must be between 200 and 240 MeV, and the pressure $P$=0 at the
saturation density $\rho_0$. The optimal parameter sets we find
have $K_\infty$ around 210 MeV. As for $\rho_0$ it is about
0.18$fm^{-3}$ in the BHF calculation with three-body
force\cite{tbf}, a value somewhat larger than that usually adopted
for Skyrme parametrizations. The consequence will be a general
underestimate of radii in finite nuclei as we shall see in
Sec.IV~.

For each ($t_0,t_3$) couple previously determined we add the
constraints of fitting $\frac{E}{A}(\beta \ne 0,\rho)$ as well as
the symmetry energy $a_s(\rho)$ of infinite matter calculated in
BHF. In terms of Skyrme parameters the symmetry energy $a_s(\rho)$
is:
\begin{eqnarray}
a_s(\rho) & \equiv &
\frac{1}{2}\frac{\partial^2(E/A)}{\partial\beta^2}|_{\beta=0}
\nonumber \\
 & = & \frac{1}{3}\frac{\hbar^2}{2m}(\frac{3\pi^2}{2})^{\frac{2}{3}}\rho^\frac{2}{3}
-\frac{1}{8}t_0(2x_0+1)\rho \nonumber \\
& &-\frac{1}{24}(\frac{3\pi^2}{2})^{\frac{2}{3}}
(3\Theta_v-2\Theta_s)\rho^{\frac{5}{3}} \nonumber \\
& &-\frac{1}{48}t_3(2x_3+1)\rho^{\sigma+1} ~. \label{eq6}
\end{eqnarray}

In practice, we choose the EOS with $\beta=0.4$ as the fitting
object. In the fitting procedure we set the symmetry energy at
saturation density, $a_s(\rho_0)=34 MeV$ as a constraint for the
full EoS of asymmetric nuclear matter. Thus, we obtain several
possible solutions for ($t_0,x_0,t_3,x_3$) with $t_0, t_3$ already
given by the symmetric matter fits and $x_0,x_3$ determined by
non-symmetric matter properties. Figs.4-5 display typical fits
obtained in this way. In Fig.4 are shown the energies per particle
in symmetric and $\beta$=0.4 non-symmetric nuclear matter,
calculated in BHF and by the Skyrme procedure. The pure neutron
matter case, which is not included in the fitting procedure, is
also shown to demonstrate that the present Skyrme parametrizations
describe reasonably well the BHF energies per particle in the
whole range of $\beta$ from 0 to 1 and for densities up to the
highest BHF calculated values around 0.35 $fm^{-3}$. However, one
can see that the Skyrme energy functional is not able to reproduce
accurately the EoS of pure neutron matter which exhibits an
s-shaped behaviour. This suggests that a change of the analytical
structure of the energy functional would be necessary. In Fig.5 is
shown the fit of BHF symmetry energy. The deviations observed in
the neutron matter case in Fig.4 reflect in the fit of the
symmetry energy since the symmetry energy is calculated from the
expression $a_s(\rho)
=\frac{E}{A}(\rho,\beta=1)-\frac{E}{A}(\rho,\beta=0)$, and then a
bad fit for the neutron matter EOS induces a bad fit for $a_s$.

To summarize, the bulk properties of nuclear matter have enabled us
to determine several sets of ($t_0,x_0,t_3,x_3$) parameters and the
values of the parameter combinations $\Theta_s$ and $\Theta_v$ for a
fixed value $\sigma$=1/6 of the density dependence. To complete the
task we proceed to determine the remaining parameters
$(t_1,x_1,t_2,x_2)$ in the next subsection.

\begin{figure}[hbtp]
\includegraphics[scale=0.3]{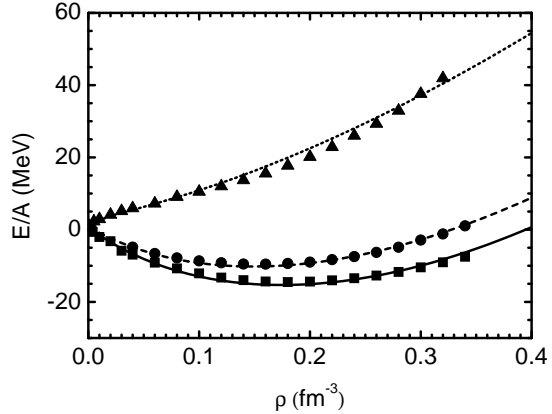}
\vglue -3.cm  \caption{Energy per particle as a function of
density. Squares, dots and triangles are BHF results for symmetric
matter, $\beta$=0.4 asymmetric matter and neutron matter,
respectively. Solid and dashed lines are results of the fit, the
dotted line is the extrapolation at $\beta$=1.} \label{Fig.4}
\end{figure}

\begin{figure}[hbtp]
\includegraphics[scale=0.3]{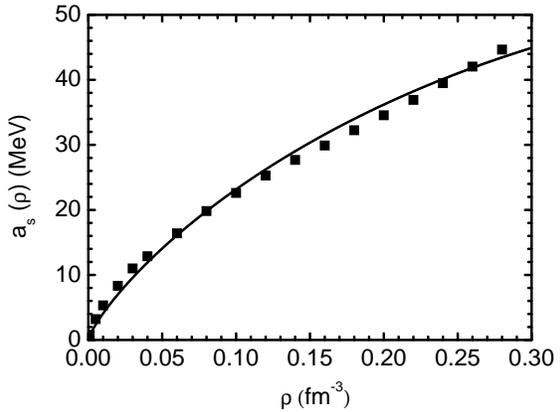}
\vglue -3.cm \caption{The symmetry energy from BHF calculations
(squares) and the present fit (solid curve).} \label{Fig.5}
\end{figure}

\subsection {Constraints from Landau parameters and finite nuclei}

\begin{center}
\begin{table}
\caption{The Skyrme parameter set and the corresponding bulk
properties of infinite nuclear matter.}
\begin{tabular}{cc}
\hline\hline
                             &    LNS        \\

$t_0(MeVfm^3)$               &  -2484.97     \\
$t_1 (MeVfm^5)$              &   266.735     \\
$t_2 (MeVfm^5)$              &  -337.135     \\
$t_3(MeV fm^{3+3\sigma})$    &   14588.2     \\
$x_0         $               &  0.06277      \\
$x_1$                        &   0.65845     \\
$x_2         $               &  -0.95382     \\
$x_3$                        &  -0.03413     \\
$\sigma$                     &   0.16667     \\
$W_0(MeVfm^5)$               &    96.00      \\
\hline

$\rho_0(fm^{-3})$            & 0.1746        \\
$E/A(MeV)$                   & -15.32        \\
$K_\infty(MeV)$              & 210.85        \\
$\frac{m^*}{m}(isoscalar)$   & 0.825         \\
$\frac{m^*}{m}(isovector)$   & 0.727         \\
$a_s(MeV)$                   & 33.4          \\
\hline\hline
\end{tabular}
\end{table}
\end{center}

The bulk nuclear matter properties give only the two parameter
combinations $\Theta_s$ and $\Theta_v$. To proceed further, we can
use some additional constraints from the Landau parameters
corresponding to our reference EBHF calculation. These Landau
parameters have been investigated in Ref.\cite{zuo2003}. The two
parameters $F_0$ and $F_0'$ are related to the compression modulus
and symmetry energy, respectively, which are already incorporated
into the preceding constraints. Then, only $G_0$ and $G_0'$ are
left as constraints, and in principle they are enough to determine
the four remaining parameters. However, we prefer to use as
constraint only one of them and keep some freedom to better
optimize the surface properties of finite nuclei. Indeed, the
surface effects cannot be determined from infinite matter
calculations and we need to adjust these surface effects by
performing Skyrme-Hartree-Fock (SHF) studies of some selected
nuclei. We will choose to fit $G_0$ whose value in
EBHF\cite{zuo2003} is 0.83 at $\rho_0$=0.18 $fm^{-3}$. The
parameter $G_0$ is expressed in the Skyrme parametrization as:
\begin{equation}
G_0 = N_0 (2A+2(3\pi^2\rho/2)^\frac{2}{3}B)
\end{equation}
where $N_0$ is the level density at the Fermi surface and
\begin{eqnarray}
A=-\frac{1}{4}t_0(\frac{1}{2}+x_0)-\frac{1}{24}t_3(\frac{1}{2}+x_3)\rho^\alpha~, \\
B=-\frac{1}{8}t_1(\frac{1}{2}-x_1)+\frac{1}{8}t_2(\frac{1}{2}+x_2)
~.
\end{eqnarray}

In the SHF energy functional an important term governing surface
properties of N=Z systems is the term\cite{Chabanat} $\alpha_s
\nabla^2\rho$ where $\alpha_s=(t_2(5+4x_2)-9t_1)/32$. To determine
the 4 parameters $(t_1,x_1,t_2,x_2)$ satisfying the fixed values of
$\Theta_s$ and $\Theta_v$, and also the spin-orbit parameter $W_0$
we adopt the following procedure. We choose as reference nuclei the
closed-shell and closed-subshell nuclei $^{16}$O, $^{40,48}$Ca,
$^{56,78}$Ni, $^{90}$Zr, $^{100,132}$Sn, $^{208}$Pb. We vary
$\alpha_s$ in a range of values similar to that of usual Skyrme
forces, and for each value we obtain a set of $(t_1,x_1,t_2,x_2)$
with which we can perform a SHF calculation of the reference nuclei.
The corresponding value of $W_0$ is obtained by adjusting the
$p1/2-p3/2$ proton splitting in $^{16}$O. In this way, we can
determine the set which gives the best overall results for binding
energies and radii in the reference nuclei.

Table I summarizes the outcome of the fit. The full parameter set is
called LNS. In the lower part of Table I are shown the main bulk
properties of nuclear matter calculated with LNS. It can be noted
that $\rho_0$ is larger and the saturation energy is slightly less
negative than the empirical saturation point (0.16 $fm^{-3}$, -16.
MeV). The consequence is that, in finite nuclei central densities
will tend to be too large and radii become systematically
underestimated.

\section{HF calculations of magic nuclei}

We now discuss the results of HF calculations of finite nuclei made
with the LNS parametrization. The parameter set of Table I is
supplemented with the two-body Coulomb force. The Coulomb exchange
contributions are treated in the Slater approximation. The center of
mass correction to the total energy is approximated in the standard
way, keeping the one-body and dropping the two-body
terms\cite{Beiner}. The HF equations are solved in the radial
coordinate space, assuming spherical symmetry.

We choose to test the LNS parametrization on the following set of
closed-shell and closed-subshell nuclei: $^{16}$O, $^{40}$Ca,
$^{48}$Ca, $^{56}$Ni, $^{78}$Ni, $^{90}$Zr, $^{100}$Sn,
$^{132}$Sn, and $^{208}$Pb. The LNS force has not been fitted on
finite nuclei and therefore, one cannot expect a good quantitative
description at the same level as purely phenomenological Skyrme
forces. It is nevertheless interesting to compare its predictions
with those of a commonly used force like the SLy4 interaction. In
Fig.6 are shown the relative deviations of charge radii (upper
panel) and energies per particle(lower panel) calculated with LNS
and SLy4. As for the binding energies, one can see that LNS is
doing reasonably well in the Ca and Ni regions but it is
underbinding somewhat $^{16}$O and overbinding the medium and
heavy nuclei, the discrepancies remaining within a 5\% limit. The
SLy4 force is doing of course much better since it is adjusted on
doubly magic nuclei. The deviations of the LNS energies can be
attributed to the incorrect saturation point of the BHF equation
of state, and also to the lack of information concerning surface
properties that one should fulfill. The charge radii of LNS
exhibit a systematic behaviour of underestimating the data by 2 to
4\%. This can be understood again as a consequence of the BHF
saturation point being shifted towards a larger density. Then, the
central density in nuclei calculated with LNS becomes larger than
what it should be and this reduces the spatial extension of
nuclear densities. This is illustrated in Fig.7 where we show the
neutron and proton distributions in $^{208}$Pb calculated with LNS
and with SLy4.

\begin{figure}[hbtp]
\includegraphics[scale=0.3]{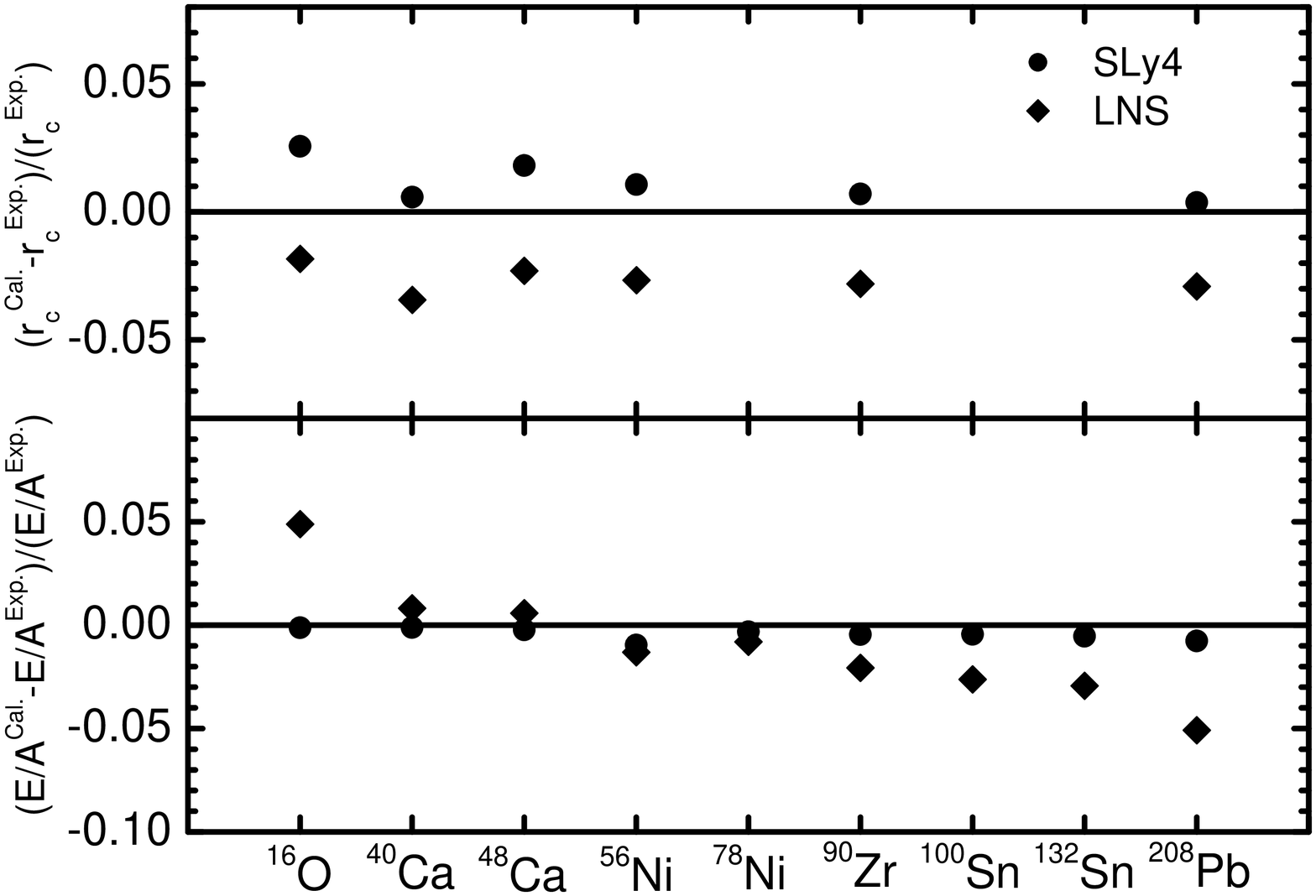}
\vglue -3.cm \caption{The relative deviations of charge radii
(upper panel) and energies per particle (lower panel) for the
selected nuclei. The results calculated with LNS and SLy4
parametrizations are represented by diamonds and circles,
respectively. } \label{Fig.6}
\end{figure}

\begin{figure}[hbtp]
\includegraphics[scale=0.3]{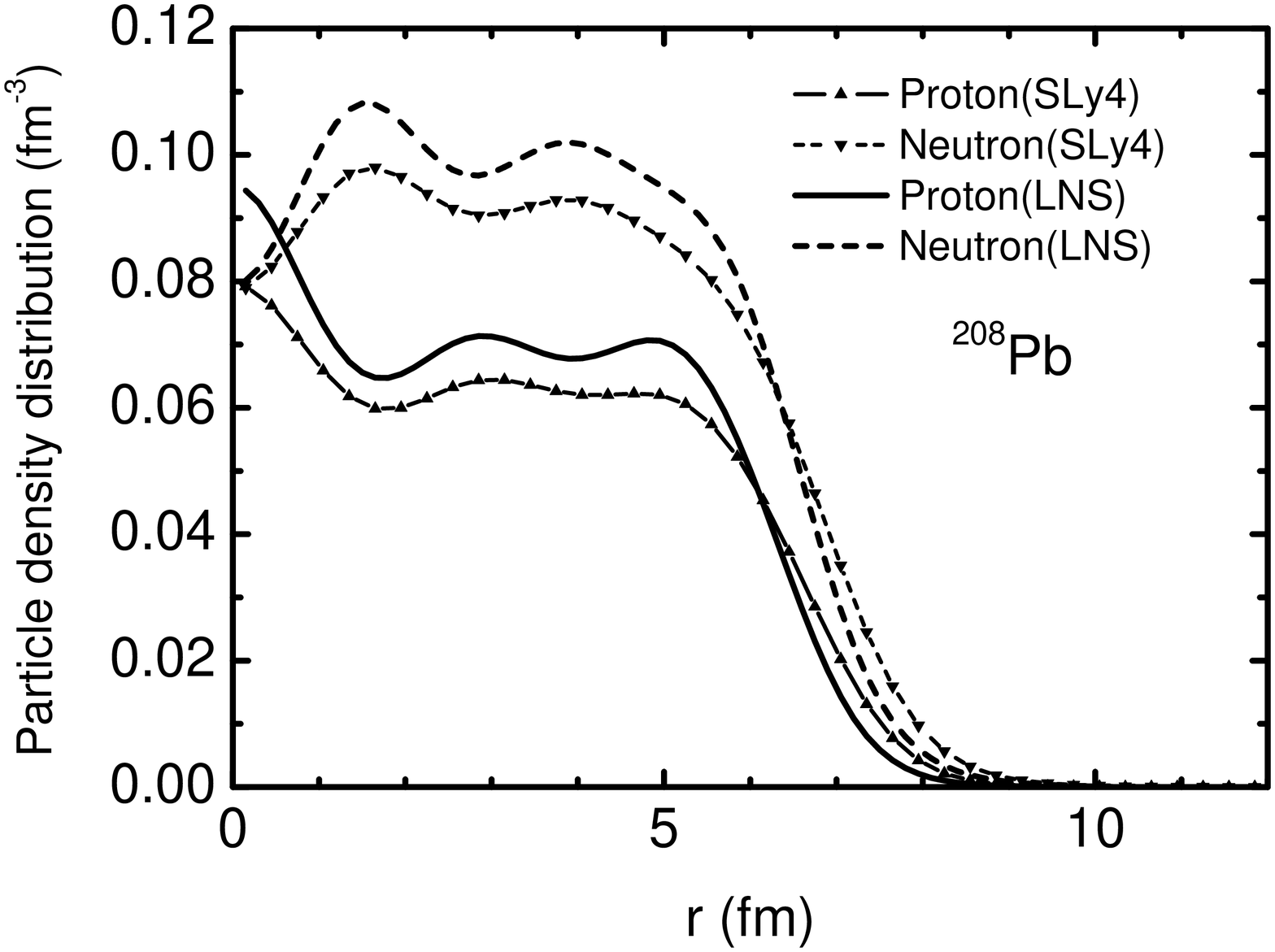}
\vglue -3.cm \caption{Neutron and proton densities in $^{208}$Pb,
calculated with LNS and SLy4 parametrizations.} \label{Fig.7}
\end{figure}

\begin{figure}[hbtp]
\includegraphics[scale=0.3]{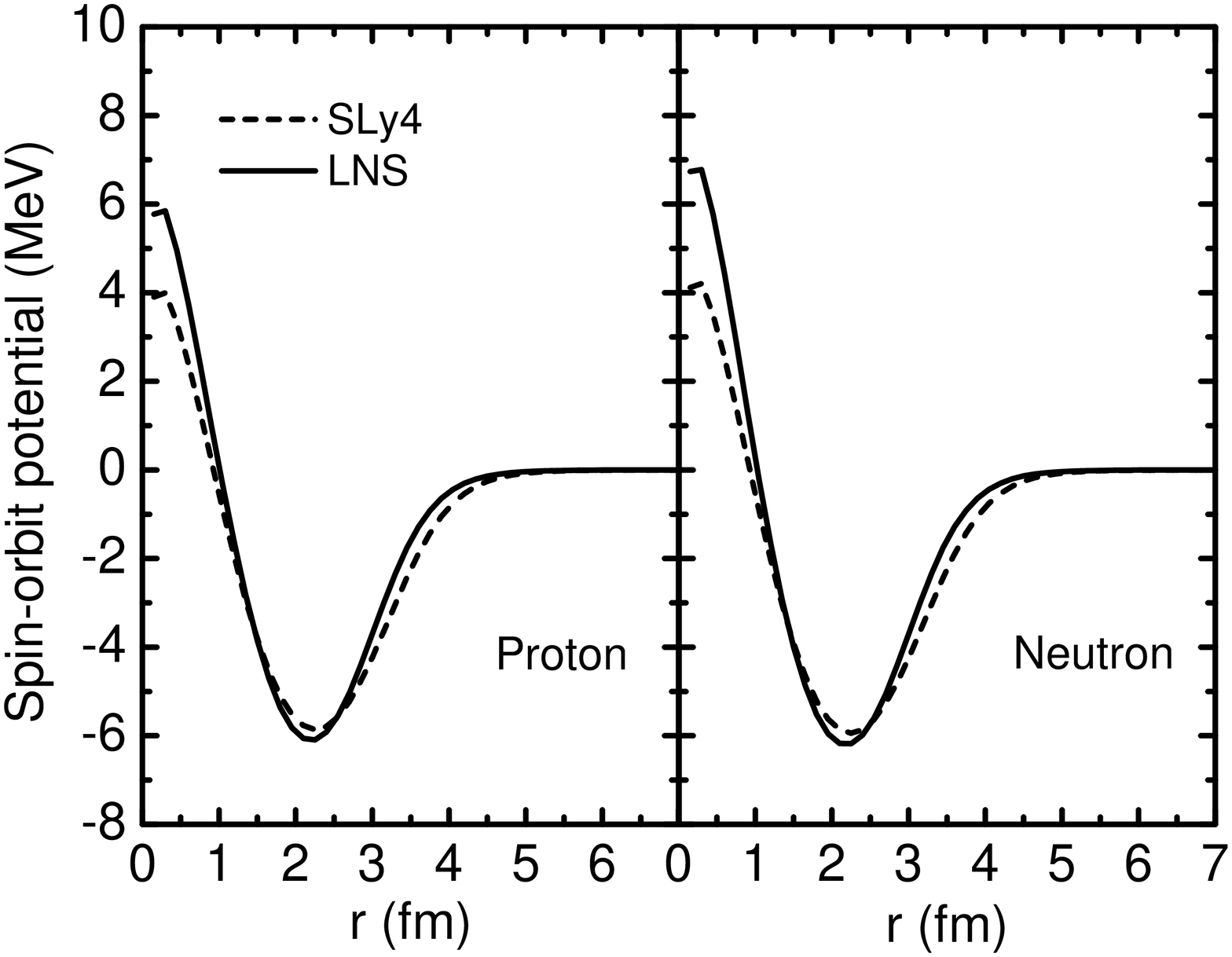}
\vglue -3.cm \caption{The spin-orbit potentials in $^{16}$O.}
\label{Fig.8}
\end{figure}

Finally, we would like to comment on the spin-orbit component of the
LNS parametrization since this is the only parameter that we could
not relate to the EBHF calculation. We have fitted it to reproduce
the experimental $1p1/2-1p3/2$ spin-orbit splitting of neutron and
proton levels in $^{16}$O. The value $W_0$=96.$MeVfm^5$ that we find
seems somewhat smaller than for other Skyrme forces where $W_0$
usually ranges from 105. to 130.$MeVfm^5$. In Fig.8 we show that the
spin-orbit potentials in $^{16}$O calculated with LNS and SLy4 are
nevertheless very close and they must give the same spin-orbit
splitting.

\section{Conclusion}
In this work we have derived a Skyrme-type parametrization of an
effective interaction suitable for Hartree-Fock calculations of
finite as well as infinite systems. The starting point is the BHF
calculations of infinite nuclear matter at different densities and
neutron-proton asymmetries. These BHF studies include effects of
three-body forces and consequently, the equation of state is much
more satisfactory than with only two body force, and the
saturation point becomes closer to the empirical point. This gives
a good motivation for looking for a simple effective interaction -
or energy density functional - whose parameters are determined by
the BHF results.

We have paid special attention to the effective mass properties in
order to get constraints on the velocity-dependent part of the
effective force. We find that, if a Skyrme force obeys the EBHF
effective mass constraints then the neutron and proton effective
masses are respectively increasing and decreasing when the asymmetry
$\beta$ becomes larger, a behaviour not always obeyed by usual
Skyrme forces, but consistent with empirical optical potential
models\cite{kon}.

The LNS force that we have thus obtained works reasonably well for
describing finite nuclei in the HF approximation. The deviations
from the data remain at the level of a few percent for binding
energies and radii. If the BHF results could be improved with
respect to the equilibrium point of the equation of state, then
the HF results of finite nuclei would most probably become much
more satisfactory. In any case, it is highly desirable to
establish a link between microscopic many-body theories which are
carried out in infinite systems and phenomenological approaches
for finite nuclei based on effective interactions. The present
work is one step into this direction.

\subsection*{Acknowledgments}

This work was performed within the European Community project {\it
Asia-Europe Link in Nuclear Physics and Astrophysics},
CN/ASIA-LINK/008(94791) .

\end{document}